\documentclass[aps, prb, twocolumn, oneside, 10pt, amsfonts, amsmath, nofootinbib, floatfix, superscriptaddress]{revtex4-2}
\usepackage{graphicx}
\usepackage[colorlinks=true, citecolor = red]{hyperref}
\usepackage{physics}
\usepackage{hhline}
\usepackage{caption}
\usepackage{subcaption}
\usepackage{blkarray}
\usepackage{kbordermatrix} 
\usepackage{enumitem}
\renewcommand{\kbrdelim}{)}
\renewcommand{\kbldelim}{(}
\usepackage{nicematrix}
\usepackage{xcolor}
\usepackage{wasysym}
\newcommand\mdoubleplus{\mathbin{+\mkern-10mu+}}
\newcommand{\Fg}{F$_1$} 
\newcommand{\Frac}[2]{F$_{#1}^{#2}$}

\begin{document}
\title{Metallic mean fractal systems and their tilings}
\author{Sam Coates}
\affiliation{Surface Science Research Centre and Department of Physics, University of Liverpool, Liverpool L69 3BX, UK}
\email{Corresponding author: samuel.coates@liverpool.ac.uk}

\begin{abstract}
Fractals and quasiperiodic structures share self-similarity as a structural property. Motivated by the link between Fibonacci fractals and quasicrystals which are scaled by the golden mean ratio $\frac{1+\sqrt{5}}{2}$, we introduce and characterize a family of metallic-mean ratio fractals. We calculate the spatial properties of this generalized family, including their boundaries, which are also fractal. We then demonstrate how these fractals can be related to aperiodic tilings, and show how we can decorate them to produce new, fractal tilings.
\end{abstract}
\date{\today}
\maketitle

\section{Introduction}

Fractals can be used to understand the behaviours of a range of complex systems, both natural and manufactured \cite{Feder2013fractals, Mandelbrot1977fractals, Barnsley2014fractals}. Their application as models for biological, chemical, and physical systems is varied and wide, and there is also particular interest in the properties of fabricated fractal materials. Within the context of physical media and/or condensed matter physics \cite{Nakayama2009fractal}, the use or exploration of fractals includes but is not limited to the study of diffusive properties \cite{Isichenko1992percolation, Fernandez2013random, O1985diffusion,O1985analytical}, topological states (electronic, photonic, acoustic etc.) \cite{Pai2019topological, Manna2022higher,Yang2020photonic, Li2022higher, Jha2023properties, Zheng2022observation}, and tunable metamaterials and antennas \cite{Miyamaru2008terahertz,Song2016broadband, Wang20233d, Tang2015fractal,Gianvittorio2002fractal, Karmakar2021fractal}.

The self-similarity of fractals is an attribute shared by many quasicrystalline materials or aperiodic tilings, and, the properties of such systems can be fractal \cite{Kamiya2018discovery, Bandres2016topological, Chen2024fractal, Yuan2000energy, Dotera17, Niizeki2007dodecagonal, Thiem2013wavefunctions, Rauzy1982nombres, Ramachandrarao2000fractal}. In particular, the 1D Fibonacci sequence is a ubiquitous aperiodic structure which is routinely used as a model for quasicrystals/aperiodic media, and its links with fractality are strong \cite{Jagannathan2021fibonacci, MaceFractal}. It can be generated by two letters $A$ and $B$, using the substitutions $A\rightarrow AB$ and $B\rightarrow A$, and, when the sequence of A and B letters are drawn with geometric Lindenmayer system (L-system) rules, it produces a fractal structure \cite{Lindenmayer1968mathematical, Monnerot2009fibonacci, Ramirez2014generalization}.

Fibonacci systems and quasicrystalline structures are also directly linked by the self-similar golden-mean $\frac{1+\sqrt{5}}{2}$ -- the ratio of $A$:$B$ letters gives $\frac{1+\sqrt{5}}{2}$, and structural elements of quasicrystalline intermetallic alloys are commonly scaled by $\frac{1+\sqrt{5}}{2}$ (e.g. atom-atom distances, surface step-edge heights etc.). However, aperiodic tilings are not restricted to golden-mean scaling; tilings can be generated which are scaled by higher-order metallic-mean ratios \cite{Nakakura2019metallic, Dotera17,Matsubara2024aperiodic,Archer22}. Here, the entire family of metallic-mean sequences are produced by expanding the Fibonacci sequence substitution rules as $A\rightarrow A^mB, B \rightarrow A$, where $m = 1$ returns the golden-mean $m = 2$ is silver, $m = 3$ is bronze etc. 

Motivated by Fibonacci word geometries \cite{Monnerot2009fibonacci, Ramirez2014generalization} and these metallic-mean structures, we introduce a family of fractals with a wide variety of customisable parameters which we generalize over all metallic-mean ratios. First, we present the L-system used to create a golden-mean fractal, calculate its properties, and demonstrate its geometric restrictions. Then, we move on to generalising this fractal over all metallic-means, repeating our analysis in the general case. Lastly, we discuss how these fractals can be decorated to give aperiodic tilings. 

\section{Fractal system}

We will start by introducing the golden-mean fractal system using a basic set of parameters, before analysing the effect of varying these same parameters. Then, we will discuss the generalisation of the system (and its properties) over all of the metallic means. Here, we will refer to a metallic mean $\varphi$ in terms of $m$: 

\begin{equation}\label{Eq:metallic means}
	\varphi_m = \frac{m+\sqrt{m^2+4}}{2}
\end{equation}

\noindent where $m=1$ is the golden mean, $m=2$ is silver, and so on. Similarly, we  will discuss the fractal systems in terms of F$_m^n$, where $n$ is the generation of the fractal, and $m$ is the metallic mean value as above. So, the first generation of the golden-mean is \Frac{1}{1}, the second is \Frac{1}{2}, and so on.

\subsection{Golden-mean}

\Fg{} is described as an L-system, with the following variables, constants, and rules:

\begin{equation}\label{Eq:rules}
	\begin{aligned}
		\text{Variables:} &\quad A, B, C, A', B', C' \\
		\text{Constants:} &\quad +, - \\
		\text{Rules:} &\quad \begin{aligned}[t]
			A \rightarrow& C - B + C' + B' - C \\
			B \rightarrow& C - B + C' \\
			C \rightarrow& A \\
			A' \rightarrow& C' + B' - C - B + C' \\
			B' \rightarrow& C' + B' - C \\
			C' \rightarrow& A'
		\end{aligned} \\
	\end{aligned}
\end{equation}

\begin{figure*}[t!]
	
	\begin{equation}\label{Eq: golden_generation}
		\textstyle
		\normalsize
		\begin{aligned}	
			\text{\Frac{1}{1}}:&\, A \\
			\text{\Frac{1}{2}}:&\, C\!-\!B\!+\!C'\!+\!B'\!-\!C \\
			\text{\Frac{1}{3}}:&\, A\!-\!C\!-\!B\!+\!C'\!+\!A'\!+\!C'\!+\!B'\!-\!C\!-\!A \\
			\text{\Frac{1}{4}}:&\, C\!-\!B\!+\!C'\!+\!B'\!-\!C\!-\!A\!-\!C\!-\!B\!+\!C'\!+\!A'\!+\!C'\!+\!B'\!-\!C\!-\!B\!+\!C'\!+\!A'\!+\!C'\!+\!B'\!-\!C\!-\!A\!-\!C\!-\!B\!+\!C'\!+\!B'\!-\!C\\
			\text{\Frac{1}{5}}:&\,A\!-\!C\!-\!B\!+\!C'\!+\!A'\!+\!C'\!+\!B'\!-\!C\!-\!A\!-\!C\!-\!B\!+\!C'\!+\!B'\!-\!C\!-\!A\!-\!C\!-\!B\!+\!C'\!+\!A'\!+\!C'\!+\!B'\!-\!C \!-\!B\!+\!C'\!+\!A'\!\\
			&+\!C'\!+\!B'\!-\!C\!-\!B\!+\!C'\!+\!A'\!+\!C'\!+\!B'\!-\!C\!-\!A\!-\!C\!-\!B\!+\!C'\!+\!B'\!-\!C\!-\!A\!-\!C\!-\!B\!+\!C'\!+\!A'\!+\!C'\!+\!B'\!-\!C\!-\!B\!\\
			&+\!C'\!+\!A'\!+\!C'\!+\!B'\!-\!C\!-\!A\!-\!C\!-\!B\!+\!C'\!+\!A'\!+\!C'\!+\!B'\!-\!C\!-\!B\!+\!C'\!+\!A'\!+\!C'\!+\!B'\!-\!C\!-\!A\!-\!C\!-\!B\!+\!C'\!+\!B'\!-\!C
		\end{aligned}
	\end{equation}
	
\end{figure*}
\begin{figure*}
	\includegraphics[width=\linewidth]{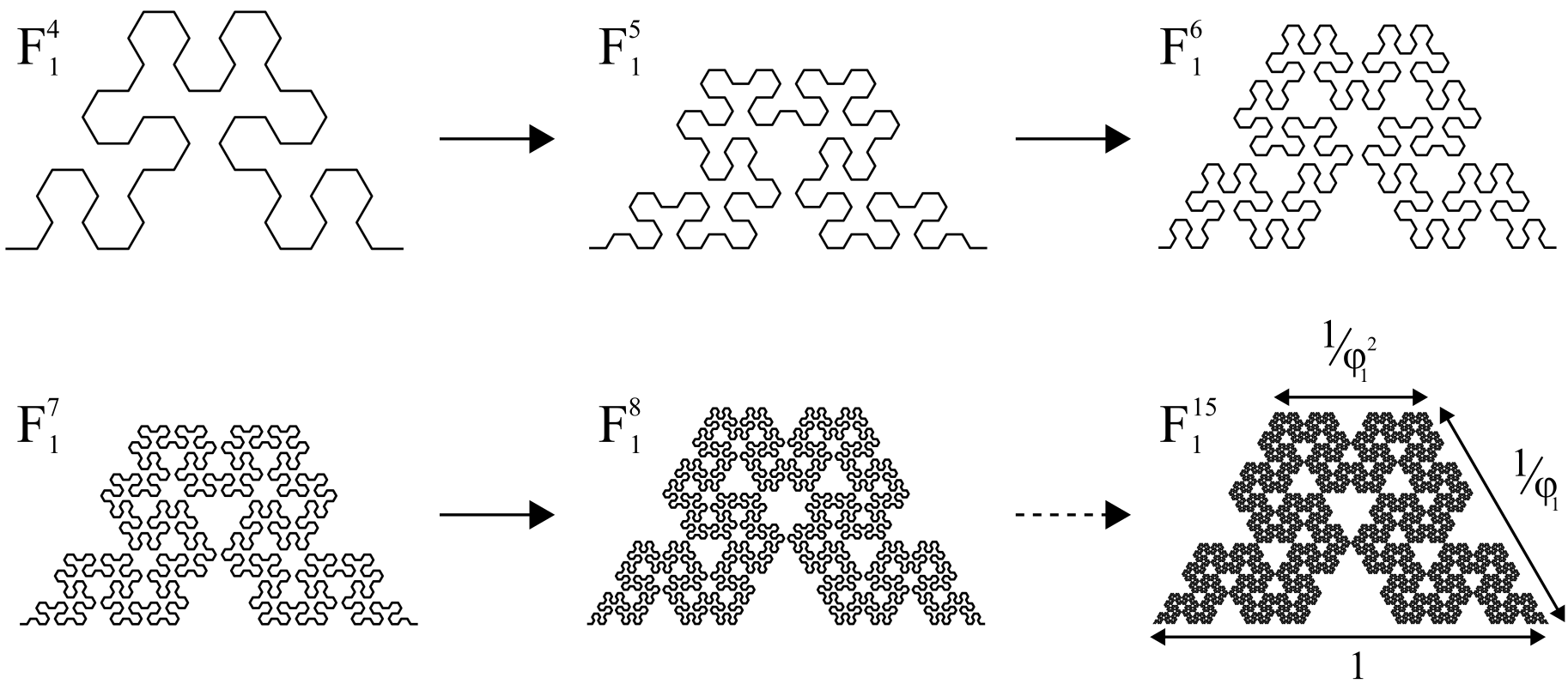}
	\caption{Generations $n = 4, 5, 6, 7, 8$ and $15$ of the \Frac{1}{} system, drawn as an L-system according to the rules of Eq. \ref{Eq:rules}, where $j=3$, and $A=B=C=1$. The scale of each generation has been normalised to 1 in $x$. Arrows highlight the approximate polygonal edges of \Frac{1}{15}, where $\varphi_1$ corresponds to Eq. \ref{Eq:metallic means} with $m=1$. \label{fig:F1}}
\end{figure*}

\noindent where variables $A$, $B$, $C$ ($A', B', C'$) are segments which move forward along a heading by an integer value, and constants $+, -$ rotate the heading by an angle $\theta$ clockwise and counter-clockwise respectively.  We note that $\theta$ is controlled by an extra parameter $j$, such that $\theta = \frac{\pi}{j}$ where $j > 2$. The rules for the $A$, $B$, and $C$ segments have mirror-symmetric counterparts in $A'$, $B'$, and $C'$: the $+$ and $-$ signs are flipped, while the rule `structure' stays the same. For the purposes of clarity, when we discuss specific properties of segments, we will refer to both $A$ and $A'$ simply as $A$ and so on. We initially set the lengths of $A = B = C = 1$ and $j=3$. We discuss the condition for $j$ and the impact of changes to $j$ and the lengths of $A, B, C$ further on. 

The first few alphabet substitution generations of \Fg{} are shown in Eq. (\ref{Eq: golden_generation}), which we initialise with an $A$ segment. We can start with any letter segment: starting with $B$ truncates the length of each generation with respect to Eq. \ref{Eq: golden_generation}, while C simply `delays' the generation by 1 (as $C\rightarrow A$). Similarly, if we start using the mirror-symmetric components, the results are identical except a simple replacement for each symmetric pair (e.g. $A:A'$, $+:-$). For simplicity, we only discuss \Fg{} as initialised from $A$ in the main text. It should be noted, however, that the properties we discuss in the following are applicable regardless of the starting segment.

Finally, as a point of interest, if we take \Frac{1}{4}, remove the $+$/$-$ constants and replace the $A$, $B$, and $C$ segments with integers 2, 1, and 0, we find a section of the ternary Fibonacci sequence: 0101020102010102010201010. This sequence takes the Fibonacci word (comprised of 1's and 0's) and replaces 00 pairs with 020 \cite{Dekking2017conjugacy, Saleh2017linear,OeisA267860}. This appears true for all even $n$ - for odd $n$, the first segment needs to be removed.

\begin{figure}
	\centering
	\includegraphics[width=\linewidth]{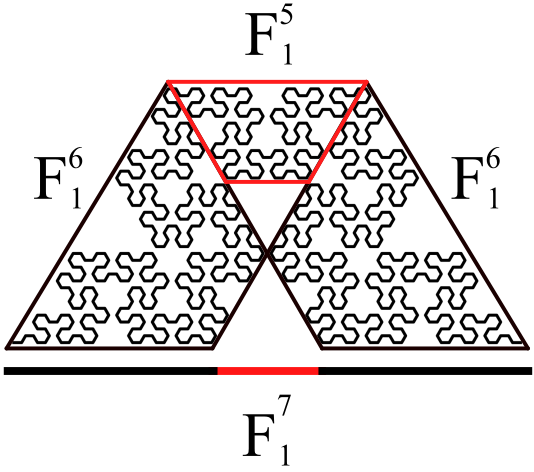}
	\caption{\Frac{1}{7} decomposed into three sub-units: two \Frac{6}{1} (black) and one \Frac{1}{5} (red). The projected contribution to the length of \Frac{1}{7} in $x$ is shown as a line, colour-coded to the sub-unit. \label{fig:segments}}
\end{figure}

\begin{figure*}[t!]
	\centering
	\includegraphics[width=\linewidth]{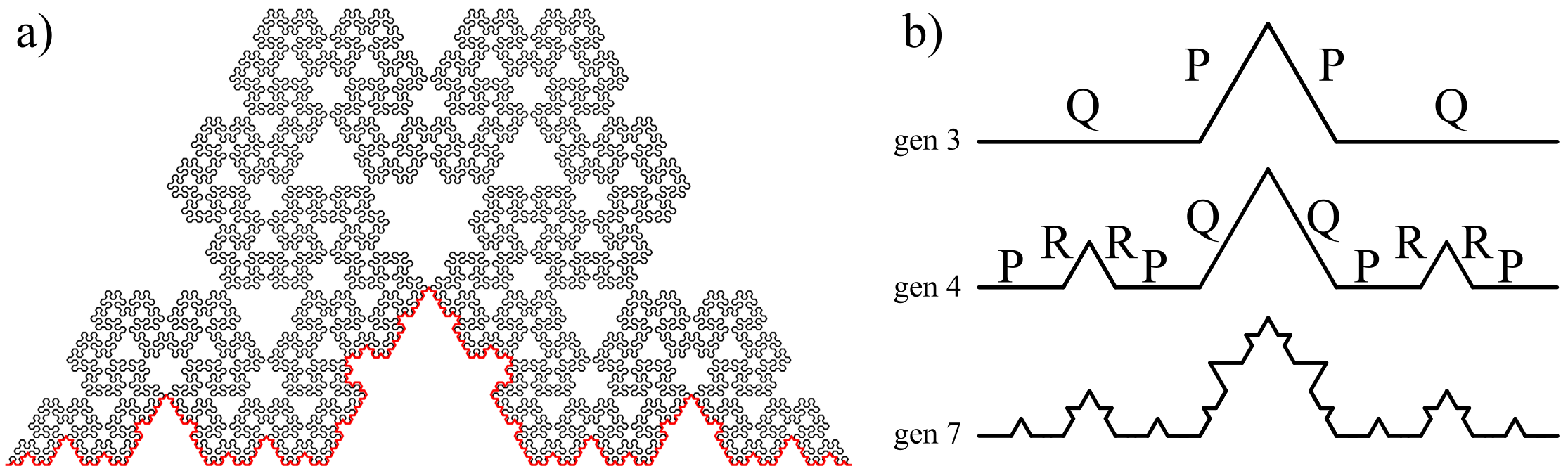}
	\caption{\textbf{(a)}\Frac{1}{10}, where the bottom boundary is overlaid and highlighted in red. \textbf{(b)} Generations of the L-system which describes the boundary of the \Frac{1}{} system. $Q, P,$ and $R$ refer to segment lengths described in the text. \label{fig:boundary}}
\end{figure*}
\subsubsection*{Geometry and the Hausdorff dimension of \Fg{}}

Figure \ref{fig:F1} shows the geometry of successive generations of \Fg{} starting from \Frac{1}{4}, with \Frac{1}{15} also shown to indicate the characteristic features of the curve at higher generations. Each curve has had its maximum scale in $x$ normalised to 1. The polygonal boundary of \Frac{1}{15} is a trapezoid with edge lengths 1, $1/\varphi_1$, and $1/\varphi_1^2$, as indicated on Figure \ref{fig:F1}. We note that initialising the sequence with a $B$ segment creates a fractal with a parallelogram polygonal boundary with edge lengths 1, $1/\varphi_1$. 

The Hausdorff dimension is a key property of a fractal, and is a measure of the scaling behaviour of a set: how the size of a set grows as its scale increases. In other words, it is the \textit{roughness} of the curve, in this case. It is calculated by \cite{Hausdorff1918dimension}:
\begin{equation}\label{Eq:hausdorff}
	H^d = \frac{\ln(N)}{\ln(S)}
\end{equation}
\noindent where $N$ represents the increase in the number of segments created in successive generations, and $S$ is the scaling factor between generations. Similar to the analysis used in the substitution rules of quasiperiodic tilings \cite{Senechal96}, inspecting the substitution matrix of the L-system rules can give us the number of segments generated after each substitution, or, $N$. Eq. (\ref{Eq:GM_segment_big}) shows the substitution matrix of \Fg{}:

\begin{equation}\label{Eq:GM_segment_big}
	M_{N} = 
	\kbordermatrix{
		& A & B & C & A' & B' & C'\\
		A & 0  & 0 & 1 & 0  & 0 & 0\\
		B & 1  & 1 & 0 & 1  & 0 & 0\\
		C &  2  &  1 & 0&  1  &  1 & 0 \\
		A' & 0  & 0 & 0 & 0  & 0 & 1\\
		B' & 1  & 0 & 0 & 1  & 1 & 0 \\
		C' &  1  &  1 & 0  &  2  &  1 & 0
	} \end{equation}

\noindent As the absolute number of segments is independent of their 'symmetry', we can simply sum the contributions of the symmetric components within each substitution rule ($A+A'$ etc.), to give the equivalent:

\begin{equation}\label{Eq:GM_segment_small}
	M_{N} = 
\kbordermatrix{
	& A & B & C\\
	A & 0  & 0 & 1 \\
	B & 2  & 1 & 0 \\
	C & 3  & 2 & 0 \\},
	\end{equation}

\noindent The largest eigenvalue, $N$, is $1+\sqrt{2}$, or $\varphi_2$, which is consistent with other Fibonacci fractals \cite{Monnerot2009fibonacci}. For completeness, the corresponding eigenvector of the eigenvalue $\varphi_2$ is proportional to (1, $\sqrt{2}$, $\varphi_2$), which tells us that the $A$ segments appear least frequently, $B$ segments $\sqrt{2}$ more often, and $C$ segments $\varphi_2$ more often than $A$ segments. 

We calculate the scaling factor $S$ by deconstructing \Fg{} after some generation $n$. Figure \ref{fig:segments} shows \Frac{1}{7}, which has been subdivided into three labelled segments, highlighted by their parallelogram boundaries: two are \Frac{1}{6} (black), one is \Frac{1}{5} (red). These segments are rotated (-)$\frac{2\pi}{3}$ and $\pi$ with respect to \Frac{1}{7}. The length of \Frac{1}{7}, $L^7$, can therefore be described by the sum of the black and red segments projected onto $x$, as indicated below \Frac{1}{7} in Figure \ref{fig:segments}. These projected lengths are determined by the rotation of the segments, such that we can generally describe:

\begin{equation}\label{Eq:scaling}
	\begin{aligned}
		L^n = & L^{n-1}\abs{\cos(-\frac{2\pi}{3})} + L^{n-1}\abs{\cos(\frac{2\pi}{3})}\\
	 + & L^{n-2} \abs{\cos(\pi)}, \\
	 \\
  = & L^{n-1} + L^{n-2}
	\end{aligned}
\end{equation}

\noindent Then, as we know that 

	\begin{equation}\label{Eq:Lnratio}
	\begin{aligned}
		S = \frac{L^n }{L^{n-1}} = \frac{L^{n-1}}{L^{n-2}}
	\end{aligned}
\end{equation}
\noindent such that,	
\begin{equation} \label{Eq:Ln-2}
	\begin{aligned}
		L^{n} = SL^{n-1},  \qquad L^{n-2} = \frac{L^{n-1}}{S}
	\end{aligned}
\end{equation}

\noindent we can therefore combine Eqs. Eq. \ref{Eq:scaling} and \ref{Eq:Ln-2} to give:
\begin{equation} \label{Eq:scalefactor}
	\begin{aligned}
		SL^{n-1} = & L^{n-1} + \frac{L^{n-1}}{S} \\
		S = & 1 + \frac{1}{S} \\
		S^2 - S = & 1
	\end{aligned}
\end{equation}

\noindent whose solution gives $S = \frac{1+\sqrt{5}}{2}$, meaning:

\begin{equation}
	H^d = \frac{\ln(\varphi_2)}{\ln(\varphi_1)} = 1.8316...
\end{equation}

\noindent which, following Monnerot-Dumaine \cite{Monnerot2009fibonacci}, we can write as:

\begin{equation}
	H^d = \frac{\ln(2 + \frac{2}{2+\frac{2}{2+\frac{2}{2+...}}})}{\ln(1 + \frac{1}{1+\frac{1}{1+\frac{1}{1+...}}})}
\end{equation}

\subsubsection*{Boundary}

The boundary of \Fg{} is also a fractal. Figure \ref{fig:boundary}(a) shows \Frac{1}{10}, where the bottom boundary is overlaid and highlighted in red. The left, right, and top boundaries are the bottoms of \Frac{1}{9}, \Frac{1}{9}, and \Frac{1}{8} respectively, as expected from Eq. \ref{Eq:scaling}. We also note that the perimeters of the internal triangular-like holes follow these boundary forms. The boundaries can be constructed as an L-system by considering three segments $Q, P,$ and $R$ and the same angular parameters $+$ and $-$. Eq. \ref{Eq:bound} shows the substitution rules:

\begin{equation}\label{Eq:bound}
		\begin{aligned}[t]
			Q \rightarrow&\enspace P - R + + R - P \\
			P \rightarrow&\enspace Q \\
			R \rightarrow&\enspace P
		\end{aligned}
\end{equation}

\noindent and Figure \ref{fig:boundary}(b) shows how the boundary develops for selected generations, where we have initialised with a $Q$ segment, and the lengths of the segments are $Q=P=1$, $R=1/\varphi_1$. As with the fractal itself, we can calculate $H^d$ by considering the number of segments in the boundary $NB$ at each generation by finding the largest eigenvalue of the substitution matrix:

\begin{equation}\label{Eq:gold_bound}
	M_{NB} = 
	\kbordermatrix{
		& Q & P & R\\
		Q & 0  & 1 & 0 \\
		P & 2  & 0 & 1 \\
		R & 2  & 0 & 0 \\},
\end{equation}

\begin{figure}
	
	\includegraphics[width=.7\columnwidth]{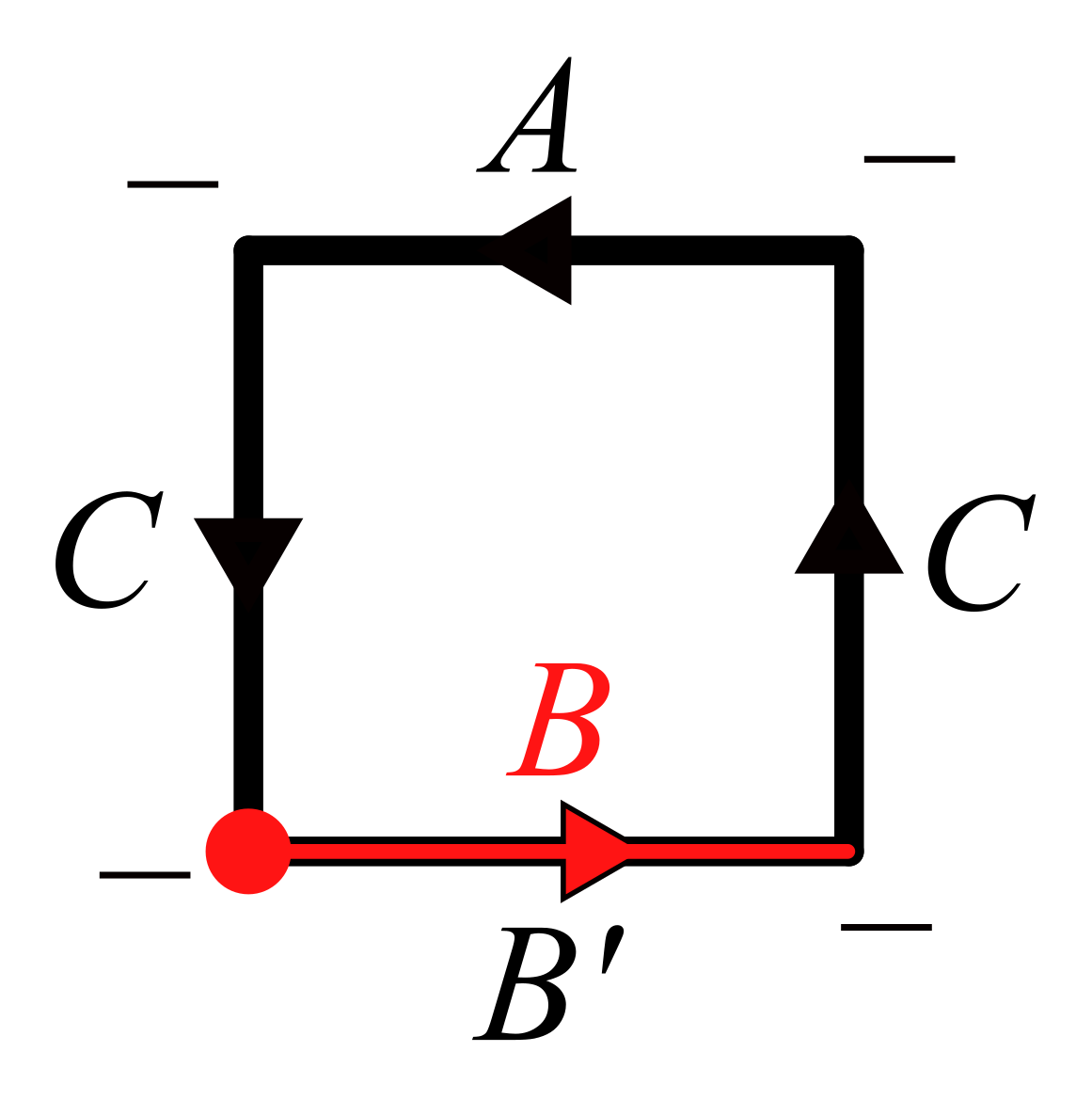}
	\caption{Schematic to show self-intersection after four successive $-$ segments when $j=2$. Sequence reads from left to right: $B'-C-A-C-B$, where the red circle indicates the end of segment $C$ intersecting with the start of $B'$, and the red arrow shows the $B$ segment being overlaid on $B'$. \label{fig:j2}}
\end{figure}

\noindent which is: 
\begin{equation*}
NB = \frac{1}{3} \left(27 - 3\sqrt{57}\right)^{\frac{1}{3}} + \frac{\left(9 + \sqrt{57}\right)^{\frac{1}{3}}}{3^{\frac{2}{3}}} = 1.7692...
\end{equation*}

\noindent Then, as $SB$ is simply the scaling factor of \Fg{}, we find 

\begin{equation}
	H^{dB} = \frac{\ln(NB)}{\ln(\varphi_1)} = 1.1857...
\end{equation}

\subsubsection*{Effect of $j$ on the Hausdorff dimension of $F^1$}

\begin{figure*}
	\includegraphics[width=\linewidth]{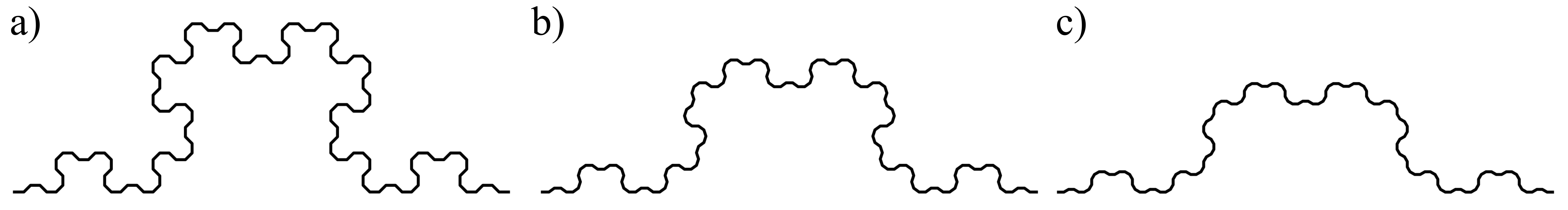}
	\caption{\textbf{(a-c)} \Fg{} for $j = 4, 5, 6$ respectively. Each curve has been normalized so that their lengths in $x$ are equal. As $j$ increases, the curve flattens, so that at the infinite limit we produce a straight line. \label{fig:increasingj}}
\end{figure*}

So far we have fixed $j = 3$. Here, we show why $j > 2$, and investigate the impact on $H^d$ when $j$ is increased. 

Trivially, when $j=1$, \Fg{} is a straight line. To show that $j\neq2$, it is convenient to analyse the distribution of the angular changes in a sufficiently long sequence. Eq. (\ref{Eq: pm_segments}) shows the $+$ and $-$ signs \textit{only} from \Frac{1}{5}:
\begin{equation}\label{Eq: pm_segments}
	\begin{aligned}
		-&-++++----++----++++--\\
		+&+++----++++--++++----\\
		+&+----++++--
	\end{aligned}	
\end{equation}

\noindent By inspecting Eq. (\ref{Eq: pm_segments}) we can see that we have, at most, four successive + or $-$ segments. This is true over arbitrary generations, and can be proven by analysing the behaviour of adjacent segments under substitution, which is demonstrated in Appendix \ref{Sec: pm proof}. The resultant effect of this behaviour can be shown by taking a sequence of \Fg{} which includes a quadruplet of angle changes: $B'-C-A-C-B$. We note that the $-$ quadruplet always appears in this letter sequence, and the $+$ in the mirror inverse ($B+C'+A'+C'+B'$). The $-$ sequence is shown in Figure \ref{fig:j2}, where the arrows indicate the direction of the sequence as read from left-to-right. If we set the initial orientation of the $B'$ segment to be 0$^\circ$, when $j$=2 ($\theta = \frac{\pi}{2}$), the sequence is in self-contact with its initial position after the second $C$ segment, as indicated by the red circle. Then, the final $B$ segment self-intersects with the sequence, as indicated by the overlaid red section and arrow. In other words, the cumulative operation of this quadruplet on the heading is equal to $0 + \frac{4\pi}{2} = 2\pi \equiv 0$, or zero displacement, such that the fractal is guaranteed to loop and self-intersect. This looping behaviour cannot occur for $j > 2$, as  $\frac{4\pi}{j} < 2\pi$ for this condition.
 \begin{table}[]
	\setlength{\tabcolsep}{7.5pt}
	\renewcommand{\arraystretch}{1.25}
	\begin{tabular}{cccc}
		$j$    & $m = 2\cos(\frac{\pi}{j})$     & $S(m)$      & $H^d$       \\
		\cline{1-4}	3    & 1 & 1.618 & 1.832 \\
		4    & 1.414 & 1.932 & 1.338 \\
		5    & 1.618 & 2.095 & 1.192 \\
		6    & 1.732 & 2.189 & 1.125 \\
		7    & 1.802 & 2.247 & 1.089 \\
		8    & 1.848 & 2.285 & 1.066 \\
		9    & 1.879 & 2.312 & 1.052 \\
		10   & 1.902 & 2.331 & 1.041 \\
		1000 &$\sim 2$ & 2.414 & $\sim 1$       
	\end{tabular}
	\caption{The $H^d$ of \Frac{1}{} tends to 1 as $j$ increases. While the number of segments $N$ is constant, the value of $S$ is calculated by Eq. \ref{Eq:metallic means}, where $m = 2\cos(\frac{\pi}{j})$. \label{tab:jvalues}}
\end{table}

Figures \ref{fig:increasingj}(a-c) show the effect of increasing $j$ for $j = 4, 5, 6$ respectively, where the scale in $x$ is normalized to 1 for each of the curves. It is immediately obvious that we obtain curves of different forms to when $j=3$. This can be related back to Eq. \ref{Eq: pm_segments}, where we see that angular changes also come in pairs i.e. $++$ or $--$. When $j=3$, both the pair and quadruplet angle changes sum to be $> \pi$ but $< 2\pi$, leading to a dense, winding form: a consequence of construction using obtuse and reflex angles. When $j=4$, the pair changes sum to equal $\frac{\pi}{2}$, and the quadruplet $\pi$, such that we obtain a `flatter' form essentially built from acute and right angles, which removes the `winding' element. This effect only intensifies as $j$ increases.

\begin{figure}
	\centering
	\includegraphics[width=\columnwidth]{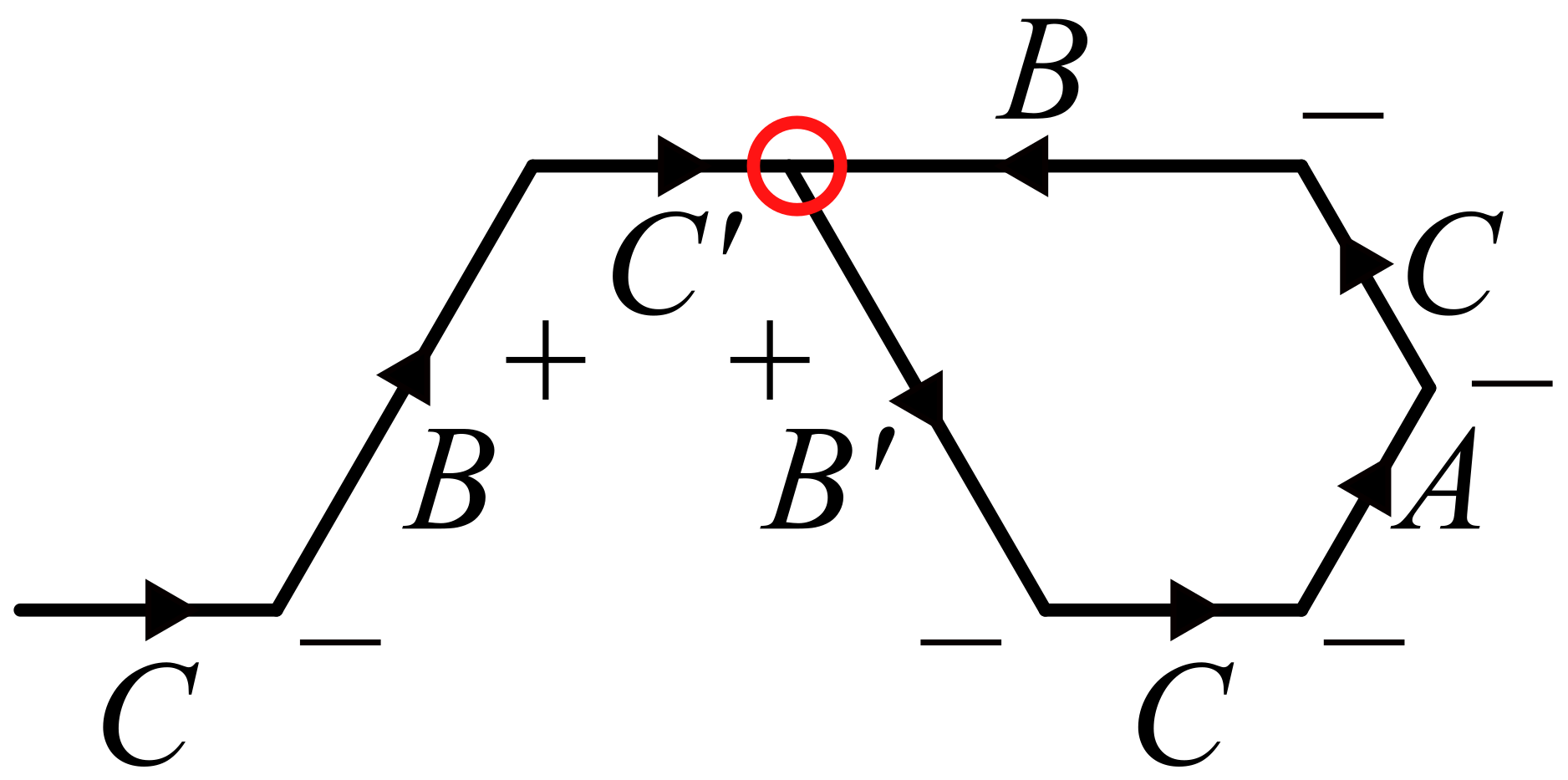}
	\caption{An example of self-contact in the \Fg{} fractal curve when $j=3$ and the length of $B$ $\geq$ $A+C$. Here, $B = 2$, and $A = C = 1$. Arrows indicate the `direction' of the fractal curve as read from left to right in the L-system. Letters and $+/-$ indicate the segments of the L-system, and a red circle indicates the point of self-contact. \label{fig:ABC}}
\end{figure}
\begin{figure*}
	\centering
	\includegraphics[width=\linewidth]{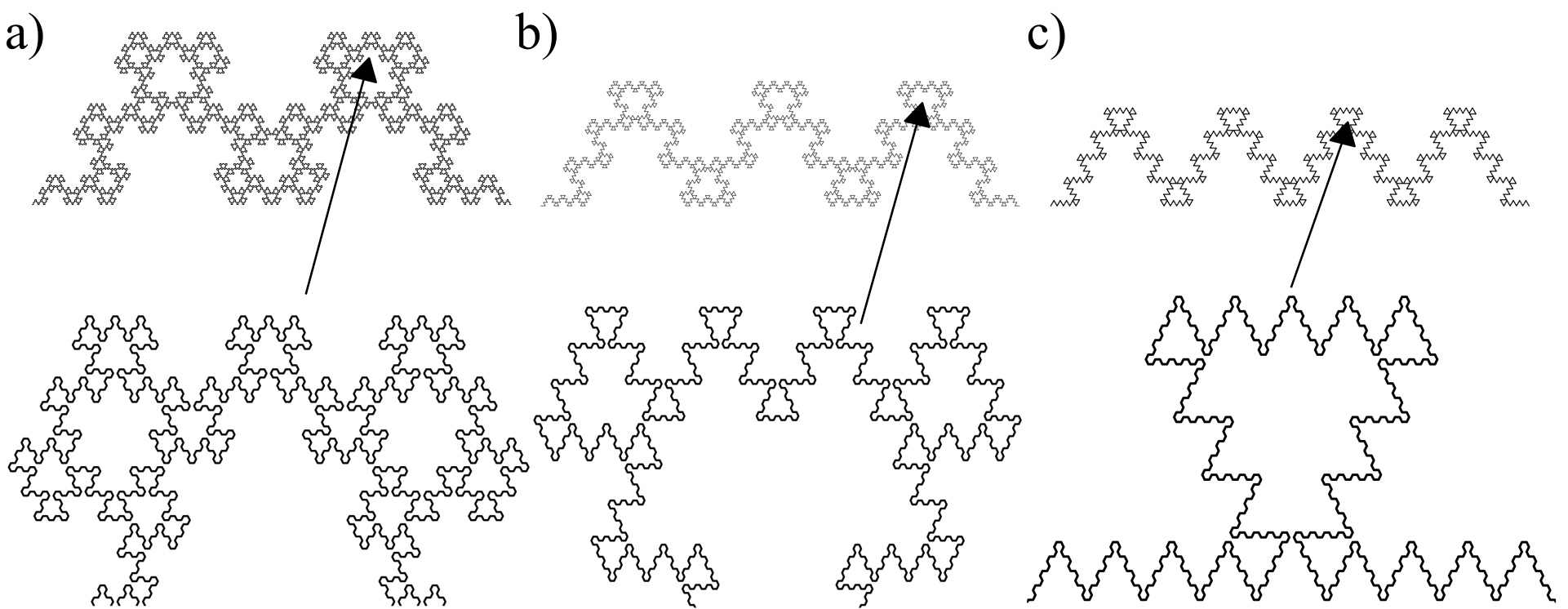}
	\caption{\textbf{(a-c)} \Frac{2}{6}, \Frac{3}{5}, and \Frac{4}{4} respectively, where $j=3$ and $A=B=C=1$. Enlarged areas correspond to those areas indicated by black arrows. \label{fig:means}}
\end{figure*}

  We now consider the change of $H^d$ of \Fg{} as $j\rightarrow\infty$; the number of segments $N$ is a constant value for each generation of $n$ regardless of $j$, meaning we only have to consider the scaling factor $S$. As the angle between segments decreases, the absolute magnitude of the length of the curve along the $x$-axis increases. This is evident when we consider that the maximum length of a single segment along the positive $x$-direction is determined by $\cos(\frac{\pi}{j})$, which is equal to $[0.5, 1]$ for $j = [3, \infty]$. So, in the infinite limit of $j$, $S$ is determined by $N$ parallel segments laid end-to-end. In other words, for $j = \infty$, $S\equiv N \equiv \varphi_2$. As such, $S$ is simply equal to Eq. (\ref{Eq:metallic means}) for $m$ values between 1 and 2, where to find $m$ for arbitrary $j$ we can use $m = 2\cos(\frac{\pi}{j})$. Table \ref{tab:jvalues} shows the values for $j, m$, and $H^d$ for a few values of $j$.

\subsubsection*{Different length values for A, B, C}

Finally, we discuss the valid values for $A, B$, and $C$, which so far have been fixed as $A=B=C=1$. The limiting factor for arbitrarily setting these lengths is self-intersection, as with our argument on the limits for $j$, this occurs when the curve `loops' during angle change segments. In other words, we need to find the conditions for the lengths of $A, B$, and $C$ which guarantee a non-zero displacement from a certain position after successive identical angle operations. We again note that we focus on systems where the lengths of the segments and their mirror-symmetric components are equal; the conditions we find likely hold when this is not the case and only require some simple extension to our proof.

We can use our analysis in the limits of $j$ to find valid lengths of $A, B$, and $C$: it is trivial that this looping can only occur when we encounter our quadruplet of angle changes. To self-intersect, and starting from an arbitrary heading, we require that the total sum of angle changes is $\geq \pi$, which can only be achieved by four angle changes when $j>2$. Similarly, if $j > 4$, the curve cannot self-intersect ($\frac{4 \pi}{j} < \pi$ for this condition). Therefore, for any $j > 4$, $A$, $B$, and $C$ can be any value so long as $A+B+C \geq 1$. We show a few examples for arbitrary parameters in Appendix \ref{Sec:arb_goldfrac}.

We can take a geometric approach for when $j=3, 4$, where we again consider the $B'-C-A-C-B$ sequence. When $j=4$, we only require that one of $A$ or $C$ are non-zero, to prevent $B$ segments directly overlapping after a cumulative heading change of $\pi$. For $j=3$, we show this sequence in Figure \ref{fig:ABC} with a few preceding segments: $C-B+C'$. Here, the lengths are $A=C=1$, and $B=2$, which leads to a self-intersection point marked in red. As stated, we require non-zero displacement from the starting point of the $B'$ segment; in other words the total displacement $\Delta > 0$, which we can write as:

\begin{equation}
	\begin{aligned}
		\Delta = &B' \cdot \begin{pmatrix} \cos(-\frac{\pi}{3}) \\ \sin(-\frac{\pi}{3}) \end{pmatrix}
		  + C\cdot \begin{pmatrix} \cos(0) \\ \sin(0) \end{pmatrix} + A\cdot \begin{pmatrix} \cos(\frac{\pi}{3}) \\ \sin(\frac{\pi}{3}) \end{pmatrix} \\
		&+ C\begin{pmatrix} \cos(\frac{2\pi}{3}) \\ \sin(\frac{2\pi}{3}) \end{pmatrix} + B\begin{pmatrix} \cos(\pi) \\ \sin(\pi) \end{pmatrix} > 0
	\end{aligned}
\end{equation}

\noindent Considering the $x$ and $y$ components separately, we find:

\begin{equation}
\Delta_x = \frac{A}{2} - \frac{B}{2} + \frac{C}{2} > 0
\end{equation}

\begin{equation}
	\Delta_y = \frac{\sqrt{3}A}{2} - \frac{\sqrt{3}B}{2} + \frac{\sqrt{3}C}{2} > 0
\end{equation}

\noindent both of which can be rearranged to give:

\begin{equation}
	A+C > B
\end{equation}
\noindent which is our condition for segment lengths when $j=3$. We show some examples for arbitrary lengths which both meet and fail this condition in Appendix \ref{Sec:arb_goldfrac}.

\subsection{Generalising over $\varphi_m$}

The fractal can be generalised over all metallic means by extending the L-system rules we defined in Eq. \ref{Eq:rules}. Using $\mdoubleplus$ to indicate the  concatenation of strings, the generalised substitution rules of the L-system can be written as:

\begin{equation}\label{Eq:gen_rules}
\begin{aligned}[t]
			A \rightarrow& C - B + C' + B' - C \mdoubleplus (m-1)\times  (-B+C'+B'-C)\\
			B \rightarrow& C - B + C' \mdoubleplus (m-1)\times (+B'-C-B+C') \\
			C \rightarrow& A \\
			A' \rightarrow& C' + B' - C - B + C' \mdoubleplus (m-1)\times  (+B'-C-B+C')\\
			B' \rightarrow& C' + B' - C \mdoubleplus (m-1)\times (-B+C'+B'-C)\\
			C' \rightarrow& A'
		\end{aligned}
\end{equation}

\noindent where $m$ is the metallic mean ratio as before, and the $\times$ operation multiplies the additional string section to be concatenated. For example, for $m=3$ and segment $A$, we would add $-B+C'+B'-C-B+C'+B'-C$ to the original rule. The conditions we have discussed for $j$ and segment lengths still hold, as these are not affected by the additional sequences added. 

\subsubsection*{Geometry and the Hausdorff dimension of \Frac{n}{}}

Figure \ref{fig:means}(a-c) shows the resultant geometry for \Frac{2}{6}, \Frac{3}{5}, and \Frac{4}{4} where $j=3$ and $A=B=C=1$. We choose these generations to more directly compare each fractal's geometry: when $m$ increases, the length of the sequences increases, so that the density of the fractal also increases for identical generations. Each of the curves has been normalised so that their length in $x$ = 1, which has the effect of `shrinking' each curve in $y$. Broadly speaking, from a macroscopic view, the curves can be described as $m$ conjoined triangular morphologies. To show the finer structure of the curves, we have enlarged areas specified by the black arrows.

To calculate $H^d$ over all $m$, we note that, trivially, $S_m$ will always be $\varphi_m$; what remains is to calculate $N_m$ for each $m$. As before, we transform Eq. \ref{Eq:gen_rules} into a substitution matrix:

 \begin{equation}\label{Eq:MM_segment_small}
	M_{N_m} = 
	\kbordermatrix{
		& A & B & C\\
		A & 0  & 0 & 1 \\
		B & 2m  & 2m-1 & 0 \\
		C & 2m+1  & 2m & 0 \\},
\end{equation}

\noindent whose largest eigenvalue for arbitrary $m$ values is:

 \begin{equation}\label{Eq:MM_eigen}
	{N_m} = \sqrt{m^2+1} +m \equiv \frac{2m+\sqrt{(2m)^2+4}}{2},
\end{equation}

\noindent meaning we can write $N_m$ = $\varphi_{2m}$, such that:

\begin{equation}
	H^d_m = \frac{\ln(\varphi_{2m})}{\ln(\varphi_m)}
\end{equation}

\noindent therefore as $m \rightarrow \infty$, $H^d_m \rightarrow 1$, or, \Frac{\infty}{n} is a straight line.

\begin{figure}
	\centering
	\includegraphics[width=.85\linewidth]{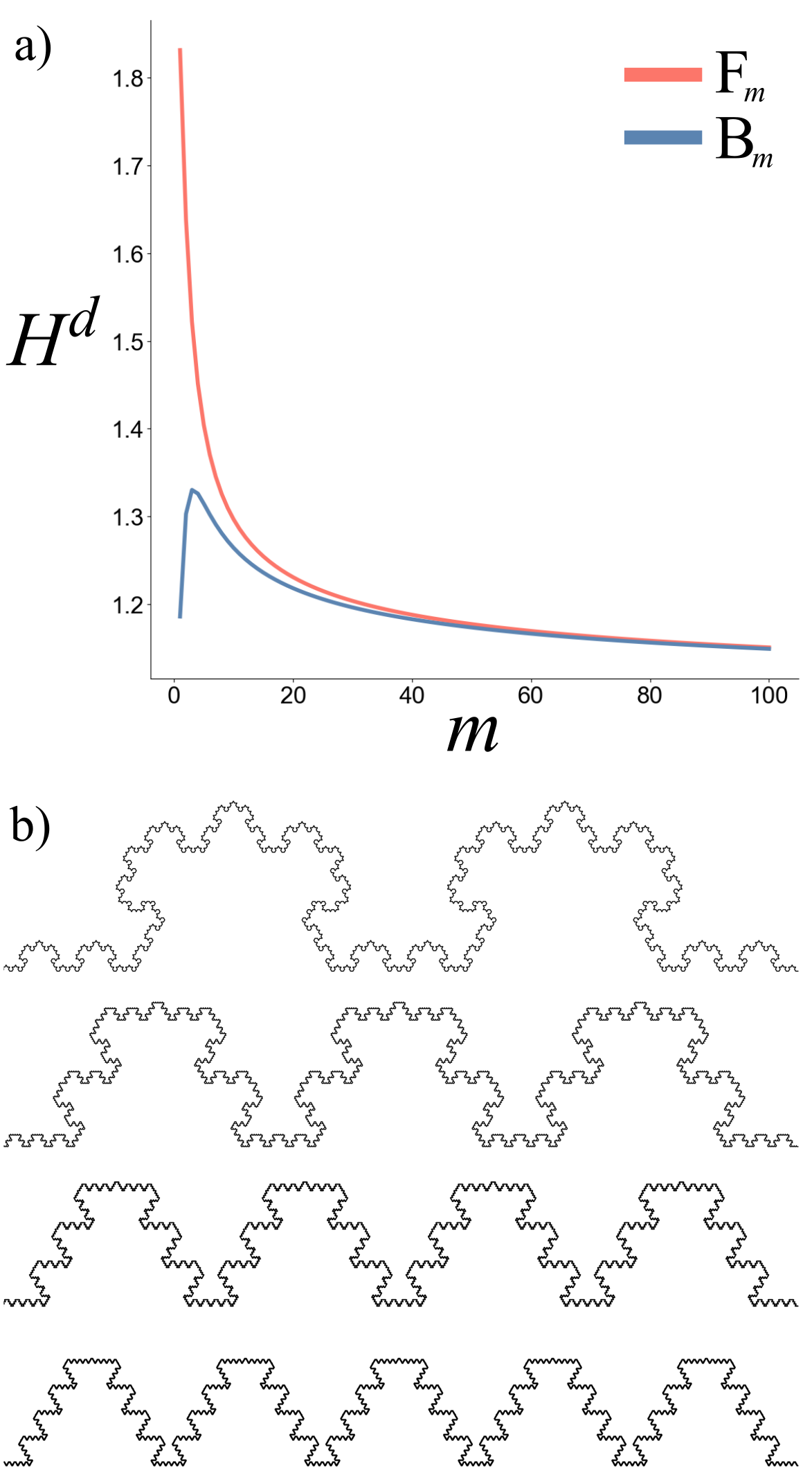}
	\caption{\textbf{(a)} The Hausdorff dimension $H^d$ for the fractals \Frac{m}{} and their boundaries B$_m$ for $m = [1,100]$. \textbf{(b)} In descending order, the boundaries of \Frac{2}{5}, \Frac{3}{5}, \Frac{4}{5}, and \Frac{5}{5}, where $j=3$ and $A=B=C=1$. \label{fig:Hd}}
\end{figure}
\subsubsection*{Boundaries}

\begin{figure*}
	\centering
	\includegraphics[width=\linewidth]{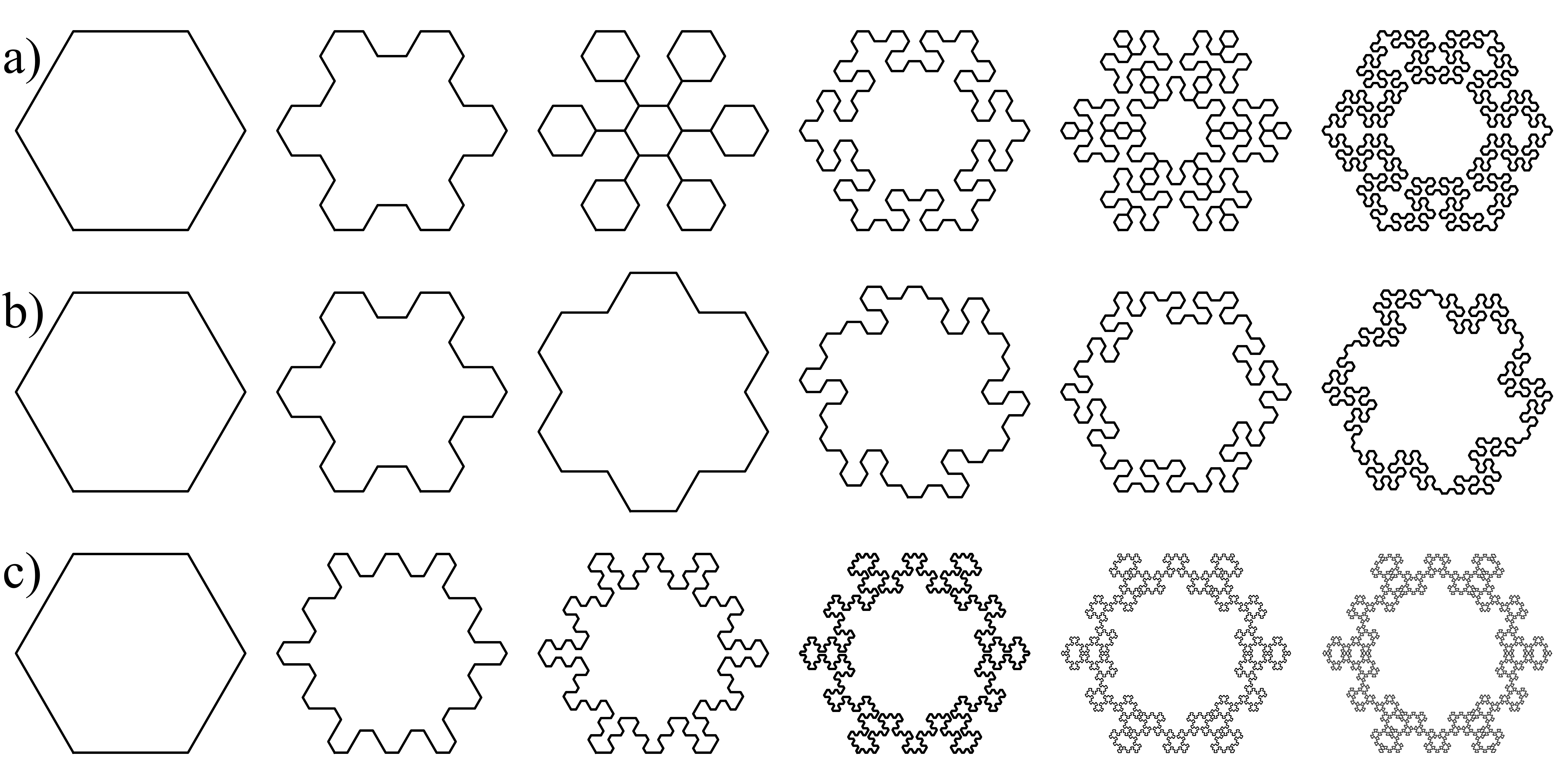}
	\caption{$j=3$ initiators and their fractals. \textbf{(a)} $A-A-$ hexagon with $m=A=B=C=1$. \textbf{(b)} $B-B-$ hexagon with $m=A=B=C=1$. \textbf{(c)} $A-A-$ hexagon with $m=2$, $A=3$, $B=2$, $C=1$.\label{fig:closedcurves}}
\end{figure*}

We can similarly extend the L-system rules which describe the \Frac{m}{} boundaries as so:

\begin{equation}\label{Eq:mean_bound}
	\begin{aligned}[t]
		Q \rightarrow&\enspace P - R + + R - P  \mdoubleplus (m-1)\times  (-R++R-P)\\
		P \rightarrow&\enspace Q\\
		R \rightarrow&\enspace P   \mdoubleplus (m-1)\times(-R++R-P)
	\end{aligned}
\end{equation}

\noindent such that we can describe the number of segments in the boundary $NB_m$ by the following matrix:

\begin{equation}\label{Eq:gen_bound}
M_{NB_m} = 
\kbordermatrix{
	& Q & P & R\\
	Q & 0  & 1 & 0 \\
	P & m+1  & 0 & m \\
	R & 2m  & 0 & 2m-2) \\},
\end{equation}

\noindent whose real eigenvalue is:

\begin{equation}
	NB_m = \frac{4m^2 - 5m + 7}{3c} + \frac{2(m - 1)+c}{3},
\end{equation}
where c is\footnote{
	\begin{equation*}
		\begin{aligned}
		c^3 = &8m^3 - 15m^2 + 3\sqrt{3}\sqrt{-m^4 + 15m^3 - 31m^2 + 45m - 9}\\
			& + 24m + 10 
		\end{aligned}
\end{equation*}}. 

As before SB$_m$ is simply $\varphi_{m}$, meaning we can also find $H^{dB}$ of the boundary for any $m$. We previously showed that as $m \rightarrow \infty$, $H^d_m\rightarrow \infty$, so it also follows that $H^{dB}_m\rightarrow \infty$ as the fractal approaches a straight line. We plot $H^d_m$ and $H^{dB}_m$ for $m = [1,100]$ in Figure \ref{fig:Hd}(a), which shows that both values tend towards both each other, and 1. Out of interest, we note that the roughness of the boundaries initially increases, and peaks at \Frac{3}{}. Figure \ref{fig:Hd}(b) shows, in descending order, the boundaries of \Frac{2}{5}, \Frac{3}{5}, \Frac{4}{5}, and \Frac{5}{5}. Upon inspection of these boundaries, it becomes clear that the branch-like structures of \Frac{2}{5} and \Frac{3}{5} gives way to a saw-tooth like curve (where of course the `teeth' also form a saw-tooth) as $m$ increases and the complexity decreases. 

\section{Tiling system}

It is possible to create a wide array of fractal geometries by initialising our L-system with any arbitrary sequence of segments. However, we are motivated by the decorations of fractal curves which give aperiodic tilings \cite{Hilbert1935stetige,Gardner1977mathematical, Pautze2021space}, and, in particular, the work on the dimer model of the Penrose and Ammann-Beenker tilings which found fractal membranes that can be decorated with the constituent tiles of rhombs (and squares) \cite{Singh2023hamiltonian, Flicker2020classical, Lloyd2022statistical}. Therefore, we focus on developing closed curve `skeletons' which can be decorated with rhombuses to form the framework of aperiodic tilings. Here, we show how to develop these skeletons using simple initiators, demonstrate the geometrical constraints for $j, A, B,$ and $C$ when decorating, and finally present and discuss some of the tilings produced.

\subsection{Skeletons and their decorations}

\begin{figure}
	\centering
	\includegraphics[width=.85\linewidth]{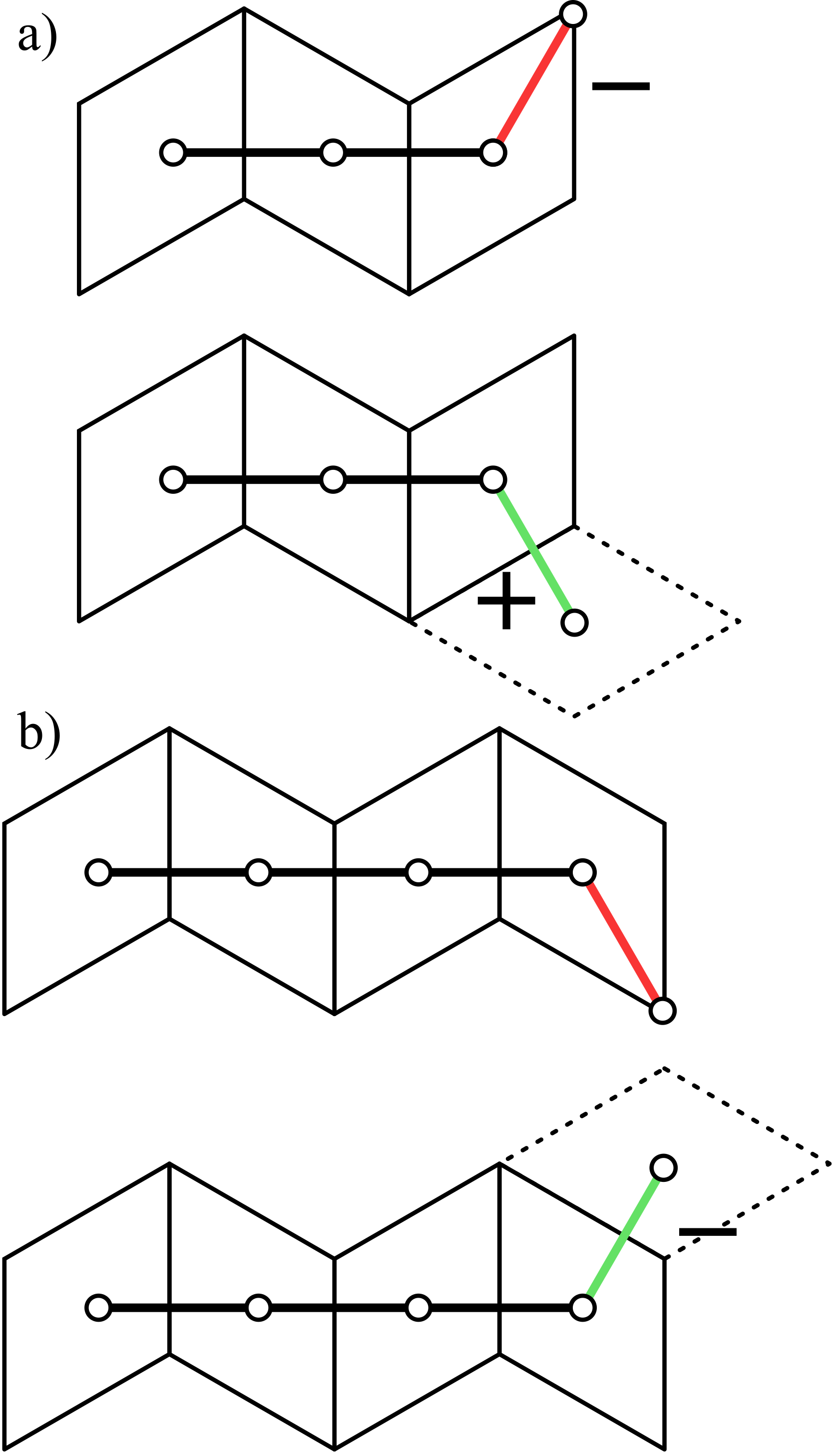}
	\caption{\textbf{(a,b)} Schematic for decorating segments of lengths 2 and 3, respectively. \textbf{(a)} shows how a $-$ angle change causes the next segment to intersect with a rhombus corner, while a $+$ segment allows a valid tile placement. \textbf{(b)} The inverse of \textbf{(a)}.\label{fig:decorate}}
\end{figure}

First, we focus on creating the simplest closed curve initiators which exhibit some rotational symmetry at their centre point. In this case we can create closed curves in terms of $j$, where all we have to do is define two `edges' of a curve and multiply this by $j$. For example, for $j=3$, we can write:

\begin{equation}\label{Eq:jseg}
	j \times (A-A-) = A-A-A-A-A-A-
\end{equation}

\noindent to create a hexagon, where we drop the superfluous $-$ segment at the end of the sequence. From here, we can apply the rules of Eq. \ref{Eq:gen_rules} to generate some fractal geometries. Figure \ref{fig:closedcurves}(a, b) show $A-A-$ and $B-B-$ hexagons where $m = A=B=C=1$. Figure \ref{fig:closedcurves}(c) shows an $A-A-$ hexagon where $m = 2$ and $A= 3, B=2, C=1$. Appendix \ref{Sec:arb_init} shows other examples with different $j$ values and arbitrary parameters.

We are motivated to find initiators and parameters which allow us to create tiling systems based on the decoration of skeletons. As such, we consider geometric arguments to find how to decorate the segments of these skeletons with respect to $j$ and the angular changes $+/-$. While it may be possible that we can decorate our skeletons with any set of polygons, we initially choose rhombus tiles whose opposing internal angles are set to either $\frac{\pi}{j}$ or $\frac{2\pi}{j}$.  We note that to fill space the fractal tilings need at least two tiles, which we discuss in the next section. In this case, we envisage that the rhombus decorations act as guides for where to place additional tiles.

Figure \ref{fig:decorate}(a, b) each show two arbitrary segments where we have set the lengths as 2 and 3 respectively, and taken $j=3$. Here, we decorate the start and end points of each unit length \textit{within} the segment, where adjacent rhombuses are mirror symmetric and we have chosen their initial orientations arbitrarily. We also show potential next steps in the L-system as red and green lines respectively. Figure \ref{fig:decorate}(a) shows that for length 2 and an angle change of $-$, the next step coincides with a rhombus vertex, whereas a $+$ change allows a potential tile placement. The converse is true for Figure \ref{fig:decorate} (b). As such, we can deduce that segments which are succeeded by a $+$ change must have an even length, and those by a $-$ change must be odd. We note that by inspection of Eq. \ref{Eq:rules}, $-$ segments always follow an $A$ or $C$, and a $+$ always follows a $B$ (the inverse is true for the mirror symmetric components as expected). Therefore, valid decorations can be found when $A, C$ are odd, and $B$ is even, for any $m$ or $j$. Of course, the inverse condition is true if we decorate the segments with mirror flipped rhombuses about the \textit{x}-axis.

\subsection{Fractal tilings}

Despite the conditions on length parameters, $j$, and the requirement that we start with an isotropic initiator, there is still an infinite set of unique fractals and therefore tiling decorations which can be generated; there are no upper bounds or limits for any of our parameters. Therefore, to limit our scope, we restrict our investigations to tilings consisting of only two types of tiles: rhombuses with opposing internal angles set to either $\frac{\pi}{j}$ or $\frac{2\pi}{j}$ as mentioned, and polygons with all internal angles set to $\frac{2\pi}{j}$. Here, we focus on a few examples of these tilings, and discuss the natural limitations as we increase $j$.

\subsubsection*{\textit{j = 3} tilings}
\begin{figure*}
	\centering
	\includegraphics[width=\linewidth]{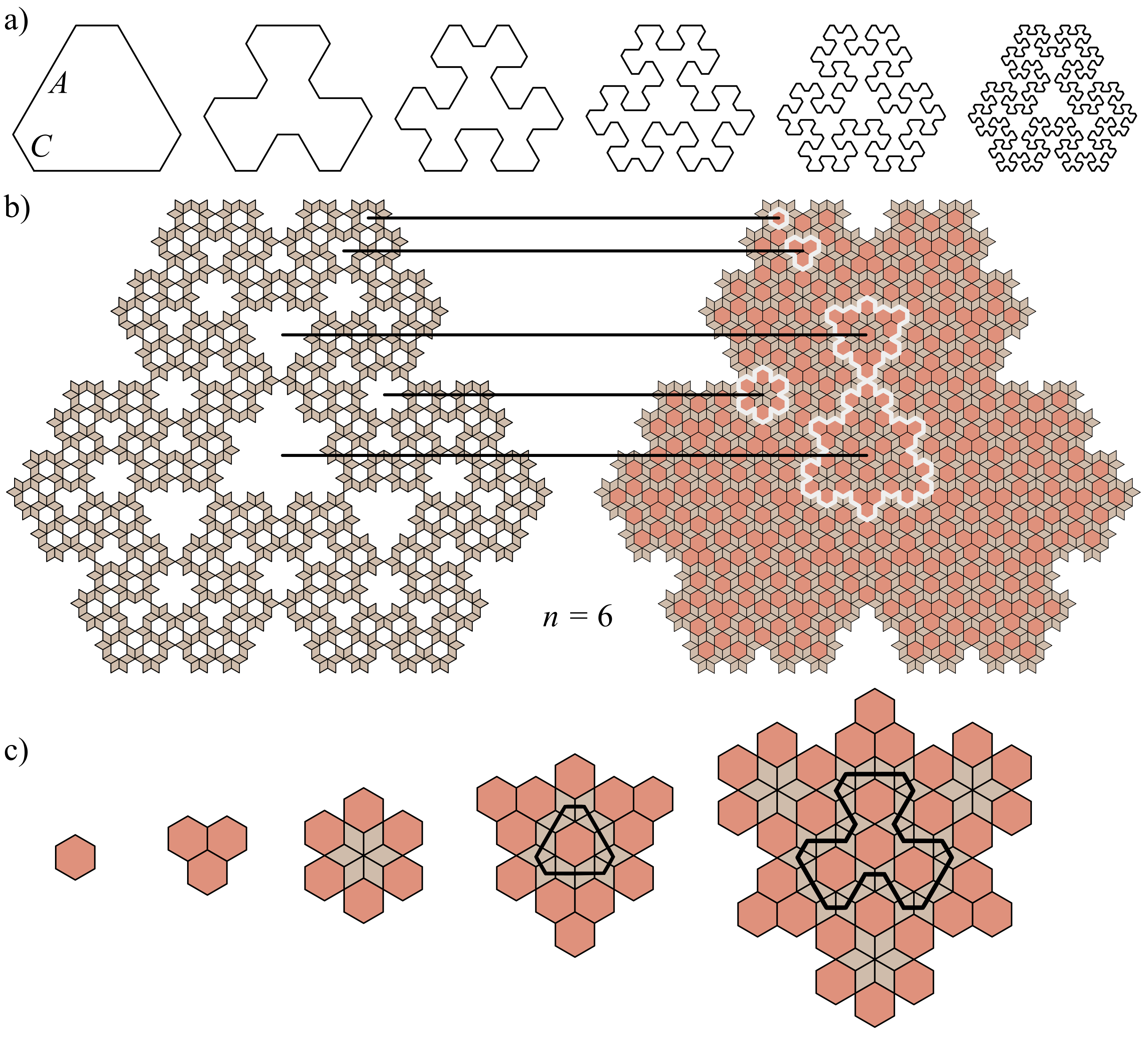}
	\caption{\textbf{(a)} An $A-C-$ initiator after 6 generations (left to right). Here, $m=1, j=3, A=3, B=2,$and $ C=1$. \textbf{(b)} Left: rhombuses decorate the sixth generation of \textbf{(a)}. Some empty spaces are indicated by black lines. Right: the empty spaces are filled by hexagonal and rhombus tiles. The black lines point to each of the corresponding spaces which are now filled, and white outlines serve to highlight the motifs. \textbf{(c)} Each of the motifs are enlarged to show their inner structure. The larger motifs appear to form early generations of the fractal form used to generate the skeleton. \label{fig:skeleton}}
\end{figure*}

To illustrate the decoration of a skeleton, Figure \ref{fig:skeleton}(a) shows the first six generations of a curve initiated by an $A-C-$ hexagon, where $m$ = 1, $A = 3, B = 2$, and $C = 1$. We have chosen these lengths as they are the first integers in the Fibonacci sequence which obey the even/odd length rules. Appendix \ref{Sec:arb_skele} shows other rhombus skeletons generated using arbitrary parameters. Figure \ref{fig:skeleton}(b), left, is the $n=6$ generation decorated by rhombs, which shows empty spaces or holes. Figure \ref{fig:skeleton}(b), right, displays how these holes can be filled by regular hexagons (internal angles of $\frac{2\pi}{j}$) and other rhombs; five of the motifs have been linked by black lines, and are also indicated by white outlines. 

\begin{figure*}
	\centering
	\includegraphics[width=\linewidth]{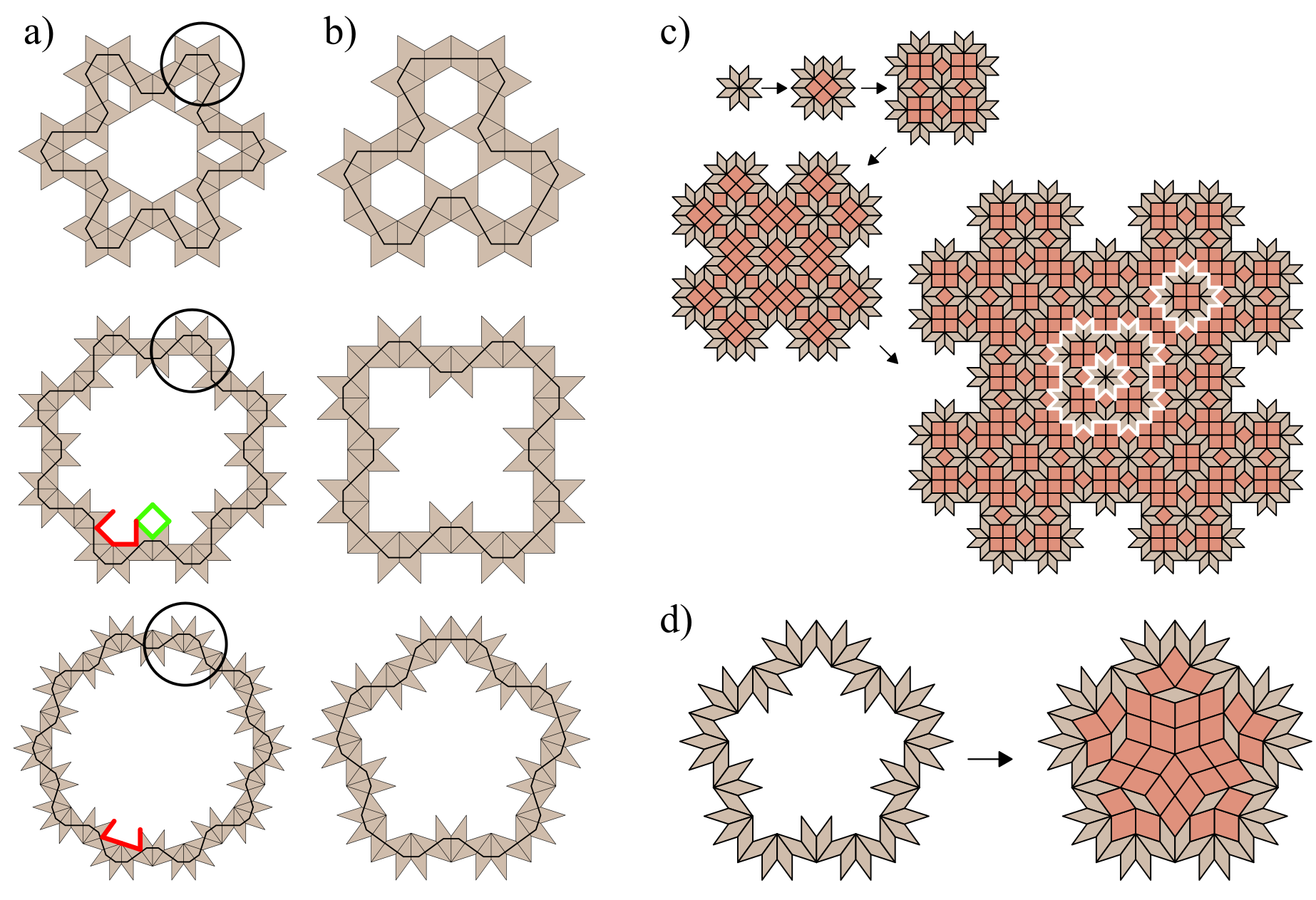}
	\caption{\textbf{(a)} Generation 3 skeletons decorated with rhombuses for $j=3, 4, 5$ in descending order. Each skeleton has been initiated using a $A-A-$ polygon for their respective $j$ values. Circles indicate sections which correspond to five successive angle changes, and green (red) lines indicate valid (invalid) geometries for a two-tile system for $j=4, 5$. \textbf{(b)} Decorated generation 3 skeletons for $A-C-$ polygon initiators which allow a two-tile system. \textbf{(c)} The creation of a $j=4$ tiling based on successive generations of the fractal system. Empty spaces have been filled with squares and rhombuses, ensuring that we recreate the earlier fractals, which are indicated by white outlines. \textbf{(d)} The generation 3, $j=5$ skeleton can be filled by fat rhombuses, but not in such a way that recreates a connected set of thin rhombus fractals. \label{fig:tilings}}
\end{figure*}

Technically, any arrangement of space-filling hexagons and rhombs could be used to pack these spaces, and the examples in Appendix \ref{Sec:arb_skele} have been left blank for those interested in doing so. However, here we found that the closed curve form of the rhombus tiles matches exactly with the distribution of rhombus tiles in the $SEH_{00}$ tiling we recently introduced \cite{Coates2024hexagonal}, as demonstrated in Appendix \ref{Sec:tiling_skele}. Therefore, we used this work as a guide to fill in the missing pieces. Figure \ref{fig:skeleton}(c) shows each of the motifs we have used, in ascending order of size, where we have rotated the final motif by $\frac{2\pi}{3}$ for easier comparison. We note that, perhaps unsurprisingly, the latter two motifs contain curves of rhombus tiles which replicate the first two generations of Figure \ref{fig:skeleton}(a), indicated by the overlaid black lines. The `generation' of these connected curves within the tiling may indicate a kind of self-similar substitution system, which is still an open question with regards to the $SEH_{00}$ tiling. It may be interesting to explore whether a set of rules exist which could produce the edges of tiles as their own fractal system, as with the Penrose tiling \cite{Ramachandrarao2000fractal}.

Considering the link between the fractal and tiling system, an alternative method to investigate arbitrary $j=3$, $m$ metallic ratio fractal systems would be to generate a tiling \textit{first}, then extract the linked rhombuses. Appendix \ref{Sec:tiling_skele} shows several examples, where we use de Bruijn's dual-grid method \cite{deBruijn81, deBruijn86} to produce $m$-mean single edge-length tilings and extract some $A, B, C$ length parameters. Recent work has considered that metallic mean hexagonal tilings with multiple edge-length tiles can approach modulated honeycomb crystals as $m \rightarrow \infty$ \cite{Matsubara2024aperiodic}. In this work, domains of large hexagon tile honeycombs are bounded by parallelograms and small hexagonal tiles; in our case, these same domains are bounded by fractals of rhombuses.

\subsubsection*{\textit{j $>3$} tilings}

Here we consider fractal tilings when $j > 3$, where again we restrict ourselves to a two-tile scheme. First, we consider further geometrical conditions when constructing an initiator. Figure \ref{fig:tilings}(a) shows decorated skeletons after 3 generations which have been initiated with an $A-A-$ sequence for $m = 1, j = 3, 4, 5$, and $A = C = 1, B = 0$. For $j=3$, it is clear that for both systems, holes can be filled by rhombuses or hexagons. However, for $j=4$, while some gaps can be filled by squares (indicated by green lines), other gaps are incongruent (red lines). A similar situation is found for $j=5$. This arises from the fact that after (twice) substituting Eq. \ref{Eq:jseg}, we find more than four successive angle changes:

\begin{equation}
	A-C-B+C+A+C+\underline{B-C-A-A-C-}B...
\end{equation}

\noindent which produces the areas circled in Figure \ref{fig:tilings}(a). In other words, $A-A-$ sequences do not occur naturally in Eq. \ref{Eq:gen_rules}. While $5\times\frac{2\pi}{3}$ leaves a $\frac{2\pi}{3}$ `gap' associated with $j=3$ geometry, $5\times\frac{2\pi}{4}$ and $5\times\frac{2\pi}{5}$ do not, and the same can be found for larger values of $j$. Changing the initiator to an $A-C-$ sequence as in Figure \ref{fig:tilings}(b), however, prevents this: $A-C-$ sequences are commonly found in Eq. \ref{Eq:gen_rules}. 

Figure \ref{fig:tilings}(c) shows a possible tiling for the $j=4$, $A-C-$ initiator, where we take into consideration the previous observations of the $j=3$/$SEH_{00}$ tiling: at each generation we fill empty spaces with squares and rhombuses, taking care to connect rhombuses so that they recreate previous fractal curves -- examples are highlighted in white in the final generation. In the same way that the $j=3$ case is related to the $SEH_{00}$ tiling, it would be interesting to explore whether the $j=4$ example has any relation to existing tilings. 

Finally, Figure \ref{fig:tilings}(d) shows that we cannot continue this specific type of tiling construction when $j=5$. From a geometric standpoint, our system doesn't allow a two-tile tiling where our second tile is a polygon with all internal angles set to $\frac{2\pi}{5}$, and the same holds for any $j>4$. However, the gaps can still be filled such that we can tile the plane, as demonstrated, although we cannot replicate the closed-curve forms of connected thin rhombuses. Whether the higher order $j$ skeletons can be decorated in a more sophisticated manner -- i.e. with more than one tile \cite{Flicker2020classical, Lloyd2022statistical, Singh2023hamiltonian} -- is an open question to explore.

	\section{Conclusions}
	
We have introduced a generalized metallic-mean fractal L-system which can be tuned by a series of parameters: the metallic-mean $m$,  the symmetry $j$, and some length-scales ($A, B, C$). First, we thoroughly explored the limits of the geometric parameters and calculated the Hausdorff dimension, $H^d$, of the golden-mean fractal, as an example. We also defined and briefly explored its boundary in terms of another L-system. Then, we generalised this system over all metallic-means, by way of expanding the original L-system with additional terms. By doing so, we were able to calculate $H^d$ for both fractal and boundary for $k\rightarrow\infty$, showing that these systems approach a straight line. 

Finally, we showed that, for some conditions, we can use these fractals as a basis for decoration, with the end goal to create fractal tilings. In our restricted system, we showed the link between a fractal system and the $SEH_{00}$ tilings, demonstrated a quick example for the $j=4$ case, and showed the limitations of our decoration method as $j$ increases. 

For future work, it may be interesting to compare the experimental and theoretical states of these systems to the abundance of work on existing fractals -- particularly those that share symmetries or are similar in morphology. For instance, the $j=3$ fractal is reminiscent of the Sierpi\'{n}ski triangle \cite{Sierpinski1915courbe} or the Koch curve \cite{Koch1904courbe}. Similarly, it might be fruitful to explore whether the tuning of our various parameters can yield properties of interest for a given symmetry or metallic mean.

\vspace{1em}

\section{Acknowledgements}

This work was supported by EPSRC grant EP/X011984/1.

\bibliographystyle{abbrv}
\bibliography{references}
\setcounter{figure}{0}
\renewcommand{\thefigure}{A\arabic{figure}}
\clearpage
\onecolumngrid
\appendix
\section{Maximum number of successive +/-- segments in \Fg{}}\label{Sec: pm proof}

First, we note that Eq. \ref{Eq:rules} does not allow identical letters between + or $-$ segments, which means that an $A$ can only be preceded or succeeded by a $B$ or $C$, and so on. If we take $A \rightarrow C-\underline{B+C'+B'}-C$, we see that the middle $C'$ segment is neighboured by two $+$, and $B/B'$ segments, as underlined. This sequence of angular changes is extended to four after we substitute for each of the segments: $A-C-B\underline{+C'+A'+C'+}B'-C-A$. The right-hand side of the $B$ substitution,  $C-B\underline{+C'}$ concatenates with the $+C'+$ substitution $+A'+$, and the left-hand side of the  $B'$ substitution $\underline{C'+}B'-C$, giving four successive $+$ segments, and no more. Similar arguments can be made for other sections under substitution. 

\section{Fractals with arbitrary \textit{j, A, B}, and \textit{C} for \textit{m} = 1}\label{Sec:arb_goldfrac}

\begin{figure*}[h!]
	\centering
	\includegraphics[width=\linewidth]{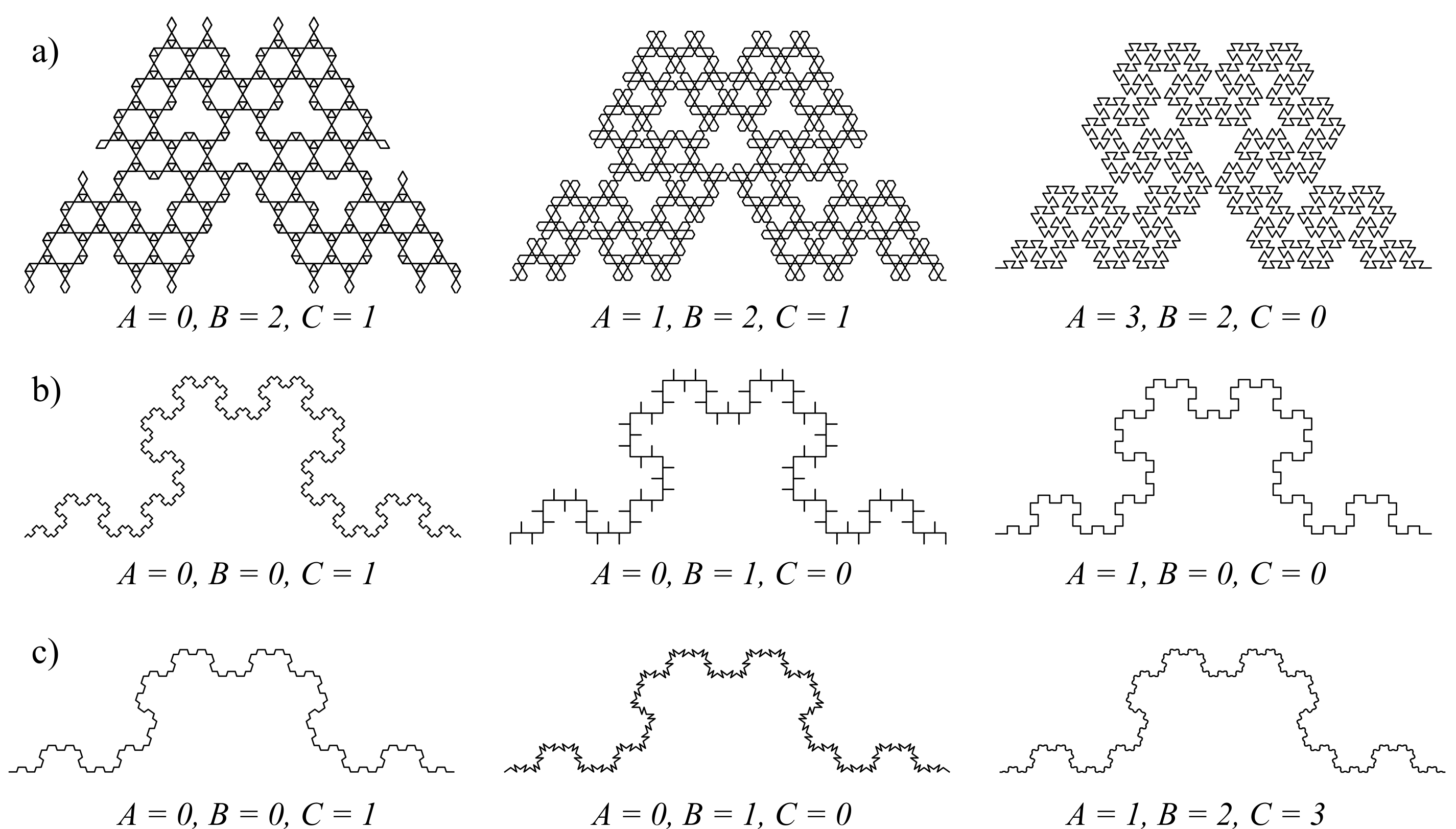}
	\caption{\textbf{(a-c)} $j = 3, 4, 5$ fractals with the length of their segments labelled. In some cases, these lengths break the conditions we set in the main text, resulting in self-intersection. \label{fig:arb_goldfrac}}
\end{figure*}

Figures \ref{fig:arb_goldfrac}(a-c) show a series of \Frac{1}{} curves for $j=3,4,5$ for arbitrary lengths. Lengths are labelled on each curve. The first two curves in Figure \ref{fig:arb_goldfrac}(a), and the second in (b) break the length conditions we discuss in the main text.

\clearpage
\section{Arbitrary initiators}\label{Sec:arb_init}

\begin{figure*}[h!]
	\centering
	\includegraphics[width=\linewidth]{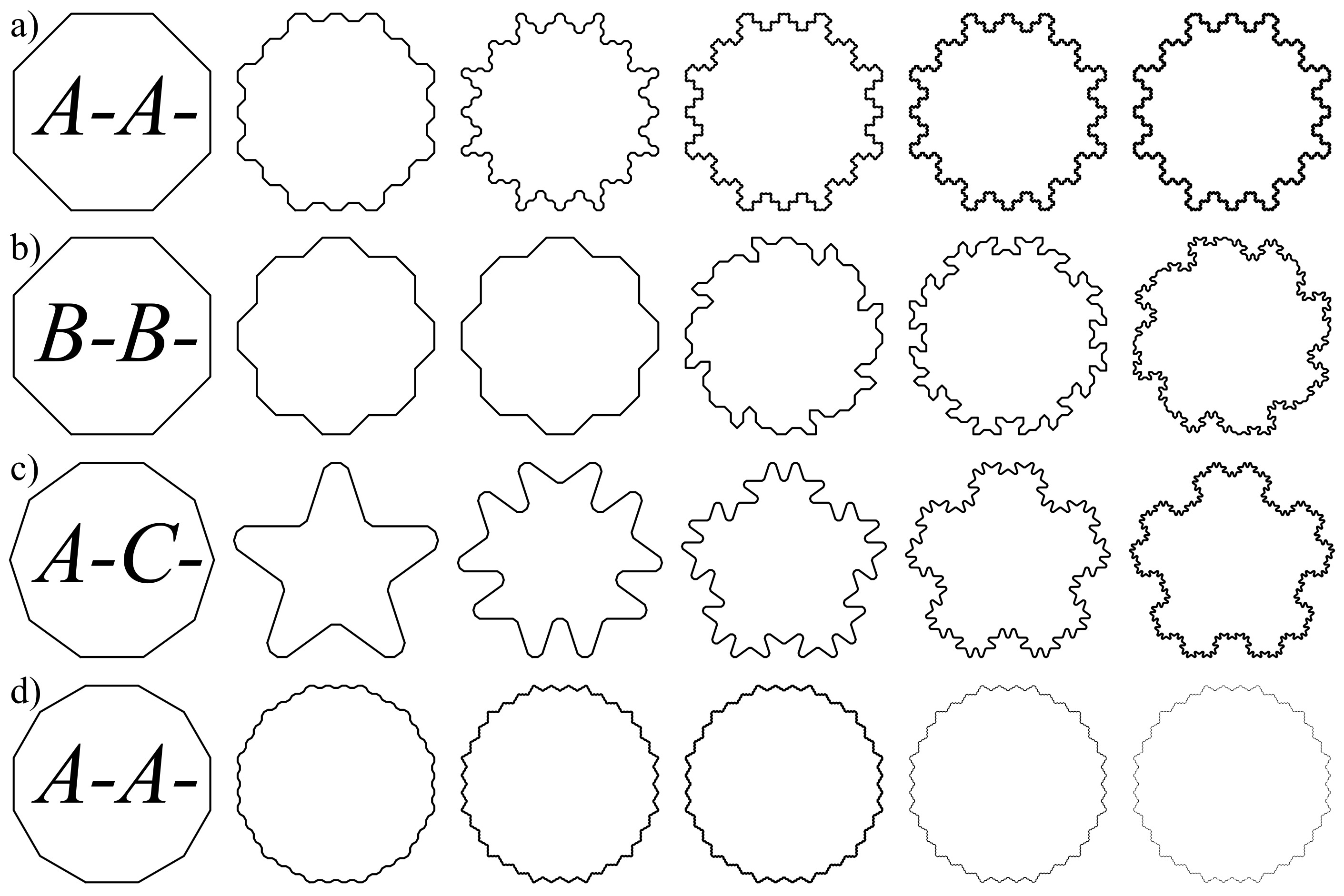}
	\caption{Arbitrary initiators, where the initial sequence used to create the first generation polygons are labelled, and the parameters are: \textbf{(a)} $m = 2, j = 4, A=1, B=1, C=1$.  \textbf{(b)} $m = 1, j = 4, A=0, B=1, C=1$. \textbf{(c)} $m = 1, j = 5, A=1, B=6, C=1$. \textbf{(d)} $m = 3, j = 6, A=1, B=1, C=1$. \label{fig:arb_init}}
\end{figure*}

Figure \ref{fig:arb_init} shows initiators for arbitrary parameters. The sequences used to generate the initiators are labelled inside the first generation, and the parameters used for the curves can be found in the figure caption.

\clearpage
\section{Skeleton decoration with arbitrary \textit{m, A, B}, and \textit{C} for \textit{j} = 3}\label{Sec:arb_skele}

\begin{figure*}[h!]
	\centering
	\includegraphics[width=\linewidth]{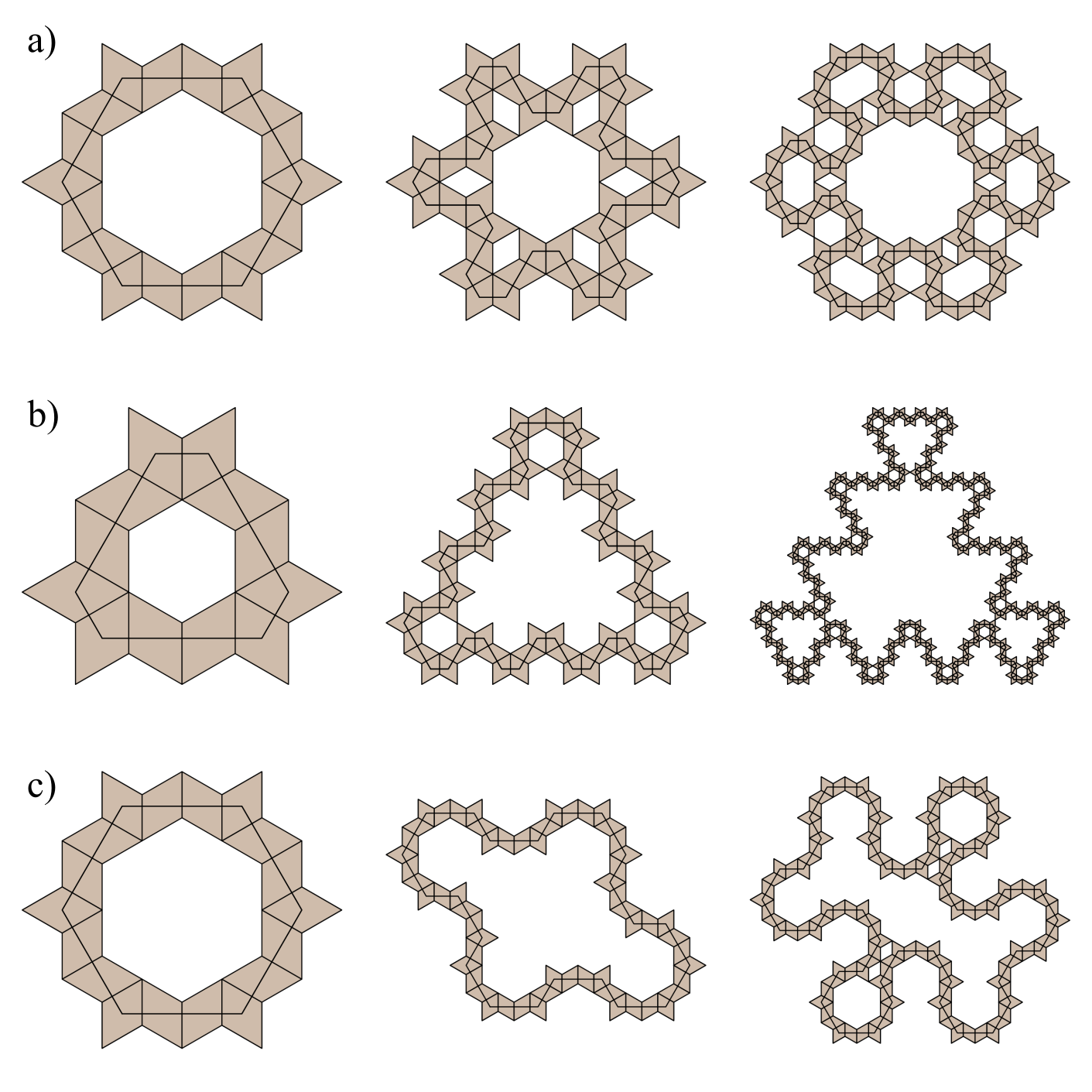}
	\caption{\textbf{(a)} $A-A-$ initiator, $m = 1, A=3, B=2, C=1$. \textbf{(b)} $A-C-$ initiator, $m =3, A=3, B=2, C=1$. \textbf{(c)} $A-C-A-A-C-A$ initiator, $m = 1, A=3, B=2, C=3$. \label{fig:arb_skele}}
\end{figure*}

Figure \ref{fig:arb_skele} shows a few examples of different initiators and parameters which can be decorated with rhombus tiles for $j=3$. Parameters can be found in the figure caption.  
\clearpage
\section{Fractal and tiling relationships for \textit{j} = 3 and \textit{m}}\label{Sec:tiling_skele}
\begin{figure*}[h!]
	\centering
	\includegraphics[width=.75\linewidth]{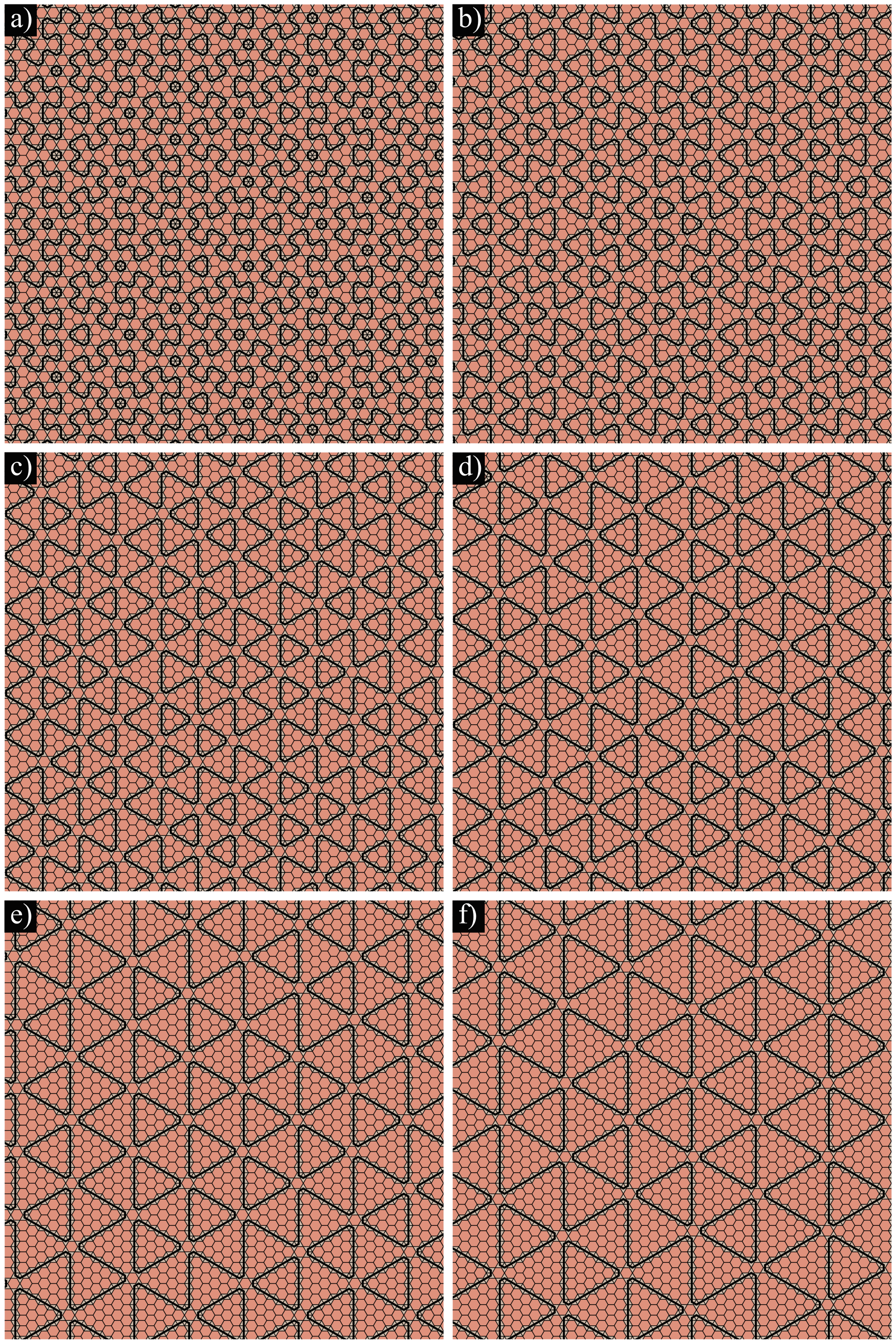}
	\caption{\textbf{(a-f)} Single edge-length hexagonal tilings generated for $m=1,...6$ respectively. Overlaid in black are the fractal curves which connect rhombus tiles. \label{fig:arb_mean}}
\end{figure*}

Figures \ref{fig:arb_mean}(a-f) show single edge-length tilings generated using de Bruijn's dual-grid method \cite{deBruijn81, deBruijn86} for $m=1,..6$. We have rotated the tilings by $30^\circ$ for graphical convenience. We overlay the fractal forms in black which are found when connecting rhombus tiles. Table \ref{tab:frac} shows the length parameters extracted from these forms after a sufficient number of generations (typically 3).

\begin{table}
		\setlength{\tabcolsep}{7.5pt}
	\renewcommand{\arraystretch}{1.25}
	\begin{tabular}{cccc}
		$m$ & $A$ & $B$ & $C$ \\
		\cline{1-4}1   &  3   &  2   &  1   \\
		2   &  5   &  4   &  1   \\
		3   &  7   &   6  &   1  \\
		4   &  9   &  8   &  1   \\
		5   &  11   &  10   &   1  \\
		6   &  13   &  12   &  1  
	\end{tabular}\caption{Length scales of the fractal forms found in metallic-mean tilings. \label{tab:frac}}

\end{table}

\end{document}